\documentclass[12pt]{article}
\usepackage{graphics}
\usepackage{epsfig}
\usepackage{amsmath}

\begin{document}

\setlength{\textheight}{240mm}
\voffset=-15mm
\baselineskip=20pt plus 2pt
\renewcommand{\arraystretch}{1.6}

\begin{center}

{\large \bf The energy of the universe in the Bianchi type-II cosmological model}\\
\vspace{5mm}
\vspace{5mm}
I-Ching Yang  \footnote{E-mail:icyang@nttu.edu.tw}

Department of Applied Science, National Taitung University, \\
Taitung 95002, Taiwan (R.O.C.)\\

\end{center}
\vspace{5mm}

\begin{center}
{\bf ABSTRACT}
\end{center}

To investigate the energy of Bianchi type-II cosmological model, I used the energy-momentum complexes 
of Einstein and M{\o}ller and obtained the zero total energy in these two prescriptions.  This result 
reinforces the viewpoint of Albrow and Tryon that the universe must have a zero net value for all  conserved 
quantities and be equivalent to the previous works of Nester et al. and Aydogdu et al.

\vspace{2cm}
\noindent
{PACS No.:04.20.Cv, 98.80.Jk.  \\}

\vspace{5mm}
\noindent

\newpage

In view of generally covariant theory, the evolution of Einstein's gravity theory, general relativity (GR), 
is underdetermined.  One problematic issue is the energy localization for the gravitational field.  In 
continuum mechanics, the most general conservation form is given by a continuity equation in a ``differential 
form" $ \partial_{\mu} J^{\mu} =0 $, and the conserved quantity would be led.  Nevertheless, the differential 
conservation law  in curved space-time will become a  covariant derivative
\begin{equation}
\nabla_{\mu} T^{\mu}_{\nu} = \frac{1}{\sqrt{-g}} \frac{\partial}{\partial x^{\mu}} 
\left( \sqrt{-g}T^{\mu}_{\nu} \right) -\frac{1}{2} 
\frac{\partial g_{\mu \alpha}}{\partial x^{\nu}} T^{\mu \alpha}=0,
\end{equation}
and does not lead to any conserved quantity.  Einstein proposed the energy-momentum pseudotensor from 
the gravitational field $t^{\mu}_{\nu}$~\cite{E1915}, follows from
\begin{equation}
\frac{1}{\sqrt{-g}} \frac{\partial}{\partial x^{\mu}} \left( \sqrt{-g}t^{\mu}_{\nu} \right)
\equiv - \frac{1}{2} \frac{\partial g_{\mu \alpha}}{\partial x^{\nu}} T^{\mu \alpha}  ,
\end{equation}
and led to energy-momentum complex
\begin{equation}
\Theta^{\mu}_{\nu} = \sqrt{-g} \left( T^{\mu}_{\nu} +t^{\mu}_{\nu} \right) ,
\end{equation}
which satisfies the differential conservation form $\partial_{\mu} \Theta^{\mu}_{\nu} = 0$.  Thus,
the conserved quantity was defined as 
\begin{equation}
P_{\nu}= \int \Theta^0_{\nu} d^3 x  .
\end{equation}
In mathematically, antisymmetric $ {\cal U}^{\mu \rho}_{\nu} $ in their two indices $\mu$ and $\rho$
would be introduced by  
\begin{equation}
\Theta^{\mu}_{\nu} \equiv \frac{\partial {\cal U}^{\mu \rho}_{\nu}}{\partial x^{\rho}}  ,
\end{equation}
and be called `` {\it superpotential} ".  There are various energy-momentum complexes which are 
pseudotensors, including those of Einstein~\cite{T62}, Tolman~\cite{T30}, Papapetrou~\cite{P48}, 
Bergmann-Thompson~\cite{BT53}, Laudau-Lifshitz~\cite{LL62}, M{\o}ller~\cite{M} and 
Weinberg~\cite{W72}.

Another idea of conserved quantity proposed by Penrose~\cite{P82} is ``{\it quasilocal} '' (i.e. associated 
with a closed 2-surface) energy-momentum according as the 4-coariant expression for the gravitational 
Hamiltonian.  Hence the conserved quantity associated with a local spacetime displacement ${\bf N}$ of 
a finite spacelike hypersurface $\Sigma$ is determined by the integral
\begin{equation}
H({\bf N}, \Sigma) = \int_{\Sigma} N^{\mu} {\cal H}_{\mu} + \oint_{\partial \Sigma} {\cal B}({\bf N}) .
\end{equation}
For any choice of ${\bf N}$ this expression defines a conserved quasilocal quanity, and there have recently 
been many quasilocal proposals~\cite{P82, K59, T83, BY93, CNT95}.
From Eq.(4), the energy-momentum within a finite region is
\begin{eqnarray}
 P({\bf N}) & = & \int_{\Sigma} N^{\nu} \sqrt{-g} \left[ T^{\mu}_{\nu} + t^{\mu}_{\nu} \right] 
d^3 \Sigma_{\mu} \\
& = & \int_{\Sigma} \left[ N^{\nu} \sqrt{-g} \left( T^{\mu}_{\nu} - \frac{1}{\kappa} G^{\mu}_{\nu} \right)  
+ \partial_{\rho} \left( N^{\nu} {\cal U}^{\mu \rho}_{\nu} \right)  \right] d^3 \Sigma_{\mu}  ,
\end{eqnarray}
where
\begin{equation}
\sqrt{-g} t^{\mu}_{\nu} = - \frac{1}{\kappa} \sqrt{-g} G^{\mu}_{\nu} + \partial_{\rho} {\cal U}^{\mu \rho}_{\nu}  
\end{equation}
and $\kappa$ is Einstein's gravitational constant. Note that ${\cal H}_{\nu}$ is the covariant form of the 
ADM Hamiltonian density, which has a vanishing numerical value.  The boundary term 2-surface integral is 
determined by the superpotential. Consequently,  Nester et al.~\cite{CNC99} show that a pseudotensor corresponds to 
a Hamiltonian boundary term
\begin{equation}
P({\bf N}) = \int_{\Sigma} N^{\nu} {\cal H}_{\nu} + \oint_{\partial \Sigma} {\cal B} ({\bf N}) 
\equiv  H({\bf N}) .
\end{equation}
Furthermore, there have been several studies aimed at uncovering the total energy of the cosmological model
by using quasilocal energy-momentum~\cite{T82,SV05,CLN07, NSV08} and energy-momentum 
pseudotensor~\cite{R94, JKSE95, BS97, X00, R00}.  In this article I would like to investigate the energy
of Bianchi type-II cosmological model using energy-momentum complex of Einstein and M{\o}ller.  
Through the paper I use geometrized units ($G=1$, $c=1$), and follow the tradition that Latin indices run 
from 1 to 3 and Greek indices run from 0 to 3.  

To describe the large scale behavior of the Universe spatially homogeneous and anisotropic cosmological 
models are used generally. Suppose that the four-dimensional spacetime manifold can be foliated by a family 
of homogeneous space-like hypersurfaces $\Sigma_t$ labeled by a constant time $t$.  Spatial homogeneity 
means that hypersurface is invariant under the three-dimensional Lie group $G_3$.  The study of $G_3$ 
led to the Bianchi classification of spatially homogeneous universes~\cite{EM69}.   The spacetime orthonormal 
coframe of cosmological model has the form $ \theta^0 = dt $ and $ \theta^a  = h^a_k (t) \sigma^k $, and 
the spatially homogeneous frame will satisfy 
\begin{equation}
d \sigma^k = \frac{1}{2} C^k_{ij} \sigma^i \wedge \sigma^j  ,
\end{equation}
where the $C^k_{ij}$ are certain constants.  There are nine Bianchi type distinguished by the particular form 
of the structure constants $C^k_{ij}$.  They fall into two special classes: class A (Types ${\rm I, II, VI_0, VII_0, 
VIII, IX}$) and class B (type ${\rm III, IV, V, VI_h, VII_h}$).  Here Bianchi type-II cosmological 
model~\cite{RB97} is considered as 
\begin{equation}
ds^2 = \eta_{\mu \nu} \theta^{\mu} \otimes \theta^{\nu} ,  \quad   \eta_{\mu \nu} = {\rm diag}  (1, -1, -1, -1) 
\end{equation}
and its spacetime orthonormal coframe is 
\begin{equation}
\begin{array}{l}
\theta^0  = dt,  \\
\theta^1  = A dx,  \\
\theta^2  = B (dy - xdz),  \\
\theta^3  = C dz .
\end{array} 
\end{equation}
In local coordinate $x^{\mu}$ the line element Eq.(12)  can be written in the form
\begin{equation}
ds^2 =g_{\mu \nu} dx^{\mu} \otimes dx^{\nu} ,
\end{equation}
and the matrix representation of metric tensor $g_{\mu \nu}$  of Bianchi type-II cosmological model would 
be expressed as
\begin{equation}
g_{\mu \nu} = \left(
\begin{array}{cccc}
1 & 0 & 0  & 0 \\
0 & -A^2 & 0 & 0 \\
0 & 0 & -B^2 & x B^2 \\
0 & 0 & x B^2 & -x^2 B^2 -C^2 
\end{array}
\right)  ,
\end{equation}
where $ A, B$ and $C$ are function of $t$.  Here, the locally rotationally symmetric (LRS) Bianchi type-II 
cosmological model~\cite{L80} is a special case of Eq.(14) as selecting $C=A$.

At the beginning the definition of the Einstein energy-momentum complex is given as
\begin{equation}
\Theta^{\mu}_{\nu} = \frac{1}{16\pi} \frac{\partial H^{\mu \sigma}_{\nu}}{\partial x^{\sigma}}  ,
\end{equation}
with the Freud's superpotential
\begin{equation}
H^{\mu \sigma}_{\nu} = \frac{g_{\nu \rho}}{\sqrt{-g}} \frac{\partial}{\partial x^{\alpha}}
\left[ \left( -g \right) \left( g^{\mu \rho} g^{\sigma \alpha} - g^{\sigma \rho} g^{\mu \alpha} \right) \right]  .
\end{equation}
Hence four-momentum will be obtained by
\begin{equation}
P_{\nu}= \int \Theta^0_{\nu} d^3 x  .
\end{equation}
The energy component of Einstein energy-momentum complex is shown as 
\begin{equation}
E_{\rm E} = \int \Theta^0_0 d^3 x  =  \frac{1}{16\pi} \int \frac{\partial H^{0i}_0}{\partial x^i} d^3 x   .
\end{equation}
According to Gauss's theorem, Eq.(18) will become a surface integral 
\begin{equation}
E_{\rm E} = \frac{1}{16\pi} \oint H_0^{\;\;0i} \cdot \hat{n}_i   dS   
\end{equation}
over the surface $dS$ with the outward normal $\hat{n}_i$.  However, all components of $H^{0i}_0$
are equal to zero.  From Eq.(19), the energy component of Einstein energy-momentum complex is
\begin{equation}
E_{\rm E} =  0  .
\end{equation}
Afterward according to the definition of the M{\o}ller energy-momentum complex~\cite{M} 
\begin{equation}
\Theta^{\mu}_{\nu} = \frac{1}{8\pi} \frac{\partial \chi^{\mu \sigma}_{\nu}}{\partial x^{\sigma}}  
\end{equation}
where the M{\o}ller's superpotential is 
\begin{equation}
\chi^{\mu \sigma}_{\nu} = \sqrt{-g} \left( \frac{\partial g_{\nu \alpha}}{\partial x^{\beta}} 
-\frac{\partial g_{\nu \beta}}{\partial x^{\alpha}} \right) g^{\mu \beta} g^{\sigma \alpha}  ,
\end{equation}
the energy component of M{\o}ller energy-momentum complex is exhibited by
\begin{equation}
E_{\rm M} = \int \Theta^0_0 d^3 x  =  \frac{1}{8\pi} \int \frac{\partial \chi^{0i}_0}{\partial x^i} d^3 x  ,
\end{equation}
As well, according to Gauss's theorem, the energy component can be written as
\begin{equation}
E_{\rm M} = \frac{1}{8\pi} \oint \chi_0^{\;\;0i} \cdot \hat{n}_i  dS .
\end{equation}
Because of no nonvanishing components of $ \chi^{0i}_0 $, the energy component of M{\o}ller 
energy-momentum complex is 
\begin{equation}
E_{\rm M} =  0  ,
\end{equation}

In conclusion, the zero total energy of Bianchi type-II cosmological models in the Einstein and M{\o}ller 
prescription have been obtained.  Albrow~\cite{A73} and Tryon~\cite{T73} supposed that the universe 
may have arisen as a quantum fluctuation of the vacuum and must have a zero net value for all conserved 
quantities.  Thus, the total energy vanishes everywhere means that the energy contributions from the material 
sources, including {\it dark matter} and {\it dark energy}, and gravitational fields inside an arbitrary two-surface
boundary $\partial \Sigma$ of the 3-hypersurface $\Sigma$ cancel each other.  Recently, Nester, So and their 
collaborator~\cite{SV05,CLN07, NSV08} indicated that the energy vanishes for all regions in Bianchi type-II 
cosmological model and my result and theirs are the same.  Furthermore Aydogdu's result~\cite{A06a,A06b} 
in locally rotationally symmetric Bianchi type-II cosmological model using Einstein and M{\o}ller 
energy-momentum complex is equivalent to the expression of my result as $C=A$.  The results of this paper 
also are consistent with those given in the previous works of Rosen~\cite{R94}, Johri et al.~\cite{JKSE95}, 
Garecki~\cite{G95} and Vargas~\cite{V04} for Friedman-Lema\^{i}tre-Robertson-Walker universe, and 
of Banerjee-Sen~\cite{BS97} and Xulu~\cite{X00} for Bianchi type-I cosmological model.  On the other 
hand, the difference of energy between the Einstein prescription $E_{\rm E}$ and the M{\o}ller 
prescription $E_{\rm M}$ is defined~\cite{Y12} as
\begin{equation}
\Delta E = E_{\rm E} - E_{\rm M} ,
\end{equation}
and its value in this article is $\Delta E =0$.  For these spacetime structures with two event horizons, like as 
stringy dyonic black hole~\cite{YCT12},  charged dilaton-axion black hole~\cite{YH14} and 
Schwarzschild-de Sitter black hole~\cite{Y17},  the difference of energy $\Delta E$ will not be equal to zero 
and is dependent on  the heat flow passing through the boundary $\partial \Sigma$.  Whatever, Eq.(27) 
would be interesting for studying the thermodynamics of cosmological model, especially the evolution of 
cosmology.

\end{document}